\documentclass{article} % For LaTeX2e
\usepackage[margin=1.2in]{geometry} 
\usepackage{times} % Kept from your original preamble
\usepackage{natbib}

\usepackage[utf8]{inputenc} 
\usepackage[T1]{fontenc}    
\usepackage{hyperref}       
\usepackage{url}            
\usepackage{booktabs}       
\usepackage{amsfonts}       
\usepackage{nicefrac}       
\usepackage{microtype}      
\usepackage{xcolor}         
\usepackage{amsmath}
\usepackage{algorithm}
\usepackage{algpseudocode}
\usepackage{graphicx}

\title{Projected Energy Matching for \\  Generative 3D Priors}

\author{%
  Daniel Barco$^\dagger$, 
  Michal Balcerak$^\ddagger$, 
  Suprosanna Shit$^\ddagger$, 
  Chinmay Prabhakar$^\ddagger$, \\
  \textbf{Philipp Denzel}$^\dagger$, 
  \textbf{Bjoern Menze}$^\ddagger$, 
  \textbf{Frank-Peter Schilling}$^\dagger$ \\
  \texttt{baoc@zhaw.ch} \\
  $^\dagger$Zurich University of Applied Sciences (ZHAW) \quad
  $^\ddagger$University of Zurich (UZH)
}

\begin{document}

\maketitle

\begin{abstract}
Transport-based generative models, which learn a time-dependent vector field that moves noise to data, have become a dominant paradigm. However, these models typically do not explicitly encode the data distribution. Energy-based models (EBMs) instead represent the data distribution explicitly through a scalar energy landscape, which \textit{Energy Matching} learns by combining transport learning with contrastive refinement. Its transport objective, however, fits energy gradients to stochastic targets, whose variance degrades the training signal at scale. We introduce \textit{Projected Energy Matching}, which learns this landscape through a more stable route: we first train a time-independent transport teacher, then freeze it and fit the negative energy gradient to its predicted velocities. This projection replaces noisy transport targets with deterministic supervision, while contrastive refinement shapes the landscape near the data manifold. On CIFAR-10, gradient-noise analysis reveals a cleaner training signal, accompanied by faster convergence than Energy Matching at matched, teacher-free training budgets. In the latent space of CT volumes, our method enables 3D CT generation and achieves better FID scores than flow models. The learned scalar potential serves as a zero-shot prior for the ill-posed inverse problem of sparse-view cone-beam CT reconstruction. By making explicit energy landscapes practical at volumetric scale, this work opens a path to wider adoption of energy-based formulations, bringing their flexibility to high-dimensional generation and inverse problems.
\end{abstract}

% --- MAIN PAPER ---
\section{Introduction}
\label{sec:intro}

Generative modeling for high-dimensional 3D data faces a trade-off between \textit{sampling efficiency} and \textit{explicit energy representation}. While Transport and Diffusion Models \citep{ho_denoising_2020,albergoBuildingNormalizingFlows2023, liu_flow_2022, lipman_flow_2022, tong_improving_2023} scale remarkably well, they typically model the distribution implicitly through their time-dependent dynamics. They do not directly expose an objective landscape $\phi(\mathbf{x})$ for comparing candidate states, combining the learned prior with other objectives, and out-of-distribution scoring. Energy-Based Models (EBMs) \citep{hinton2002training, lecun2006tutorial, du2019implicit} inherently provide this scalar potential but are notoriously difficult to train \cite{song_how_2021}. Standard  Markov chain Monte Carlo (MCMC)-based training \citep{song_how_2021, song_generative_2019, schroder_energy_2023} suffers from prohibitive computational costs, training instability, and mode collapse. Consequently, practitioners often bypass MCMC entirely by employing heavily parameterized time-dependent ensembles, or cooperative architectures which train an auxiliary generator alongside the energy model for rapid sample proposal \citep{gao_learning_2020, cui_learning_2024, guo_egc_2023, zhang_flow_2024, yoon_maximum_2024}. However, relying on auxiliary networks sacrifices the elegance of a single, unified, time-independent energy model.

Recent methods, including Energy Matching \citep{balcerak_energy_2025, balcerak2026graphenergymatching}, Equilibrium Matching \citep{wang2025equilibrium}, and Beckmann Transport Models \citep{cheuk2026beckmann}, use simulation-free supervision to learn time-independent generative dynamics. In particular, Energy Matching learns a scalar potential whose negative gradient approximates a transport velocity field ($-\nabla \phi(\mathbf{x}) \approx v(\mathbf{x})$) away from the data manifold. The resulting transport guides samples toward the data distribution, providing an effective burn-in for subsequent MCMC sampling. The transport regime, however, is trained from scratch with a double-backward pass ($\nabla_\theta \nabla_{\mathbf{x}} \phi_\theta$) guided by a stochastic interpolant target. We observe that this target need not be stochastic. A velocity model, trained cheaply with first-order optimization, already carries the transport map. We then freeze this model and regress its predicted velocities onto the negative gradient of a parameterized scalar potential, $-\nabla \phi_\theta(\mathbf{x})$, replacing noisy regression targets with deterministic supervision.

\begin{figure}[t]
    \centering
    \includegraphics[width=\linewidth]{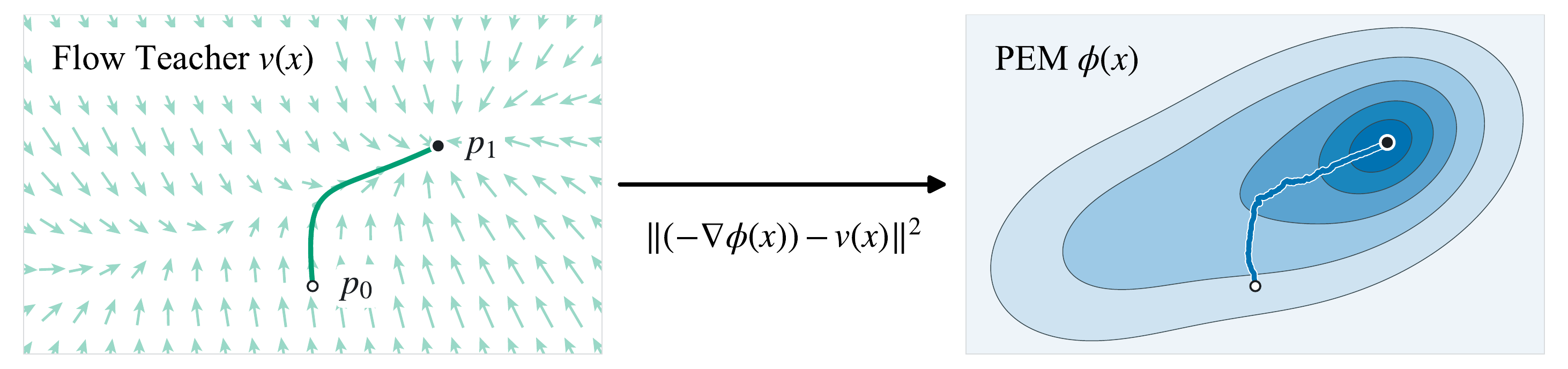}
    \caption{\textbf{Projected Energy Matching.} \emph{Left:} a pre-trained, time-independent flow teacher $v(\mathbf{x})$ (arrows) transports a sample from a noise $p_0$ to a data distribution $p_1$ along a deterministic ODE path. \emph{Right:} regressing a scalar potential onto that field under $\|v(\mathbf{x})-(-\nabla \phi_\theta(\mathbf{x}))\|^2$ projects it onto the conservative subspace of vector fields, yielding an explicit energy landscape whose gradient reproduces the teacher dynamics (sampling path, same start and endpoint).}
    \label{fig:flow2energy}
\end{figure}

Minimizing the $\ell_2$ objective between $-\nabla \phi_\theta$ and the frozen teacher field  ($v(x)$) projects the field onto the conservative subspace of vector fields as shown in Fig.~\ref{fig:flow2energy}. We term the resulting framework \textbf{Projected Energy Matching (PEM)}, which operates in three stages: (1) training a teacher model with an unconstrained flow objective to establish the transport map, (2) learning a single time-independent scalar potential by projecting the frozen teacher field onto the conservative subspace, and (3) refining the potential landscape using a contrastive loss. Both energy parameterizations of the score \citep{du_reduce_2022} and the conservativeness of learned fields \citep{chao_investigating_2023} have been studied for time-conditioned diffusion. Teacher-supervised projection has also been used to stabilize the training of time-dependent diffusion energies \citep{thornton_composition_2025}, while \citep{praneeth_velocity_2026} use a learned conservative projection to amplify a field's rotational component at inference time. In PEM, an autonomous velocity teacher supervises a time-independent transport potential, with supervision retained during contrastive refinement. The student therefore inherits the integrable part of the teacher dynamics and discards the residual.

First, we investigate whether this projected target is a better optimization problem than learning the potential from scratch. To answer it, we use gradient noise scale estimation \citep{mccandlish_empirical_2018} and measure the signal-to-noise structure of the potential-model gradient under both regimes on CIFAR-10, holding training hyper parameters fixed across experiments. Projection against a frozen teacher yields a stronger gradient signal than training from scratch, indicating that the projected target is better conditioned. This cleaner signal is the mechanism behind the compute-friendly model at 3D.

Next, we train the potential on 25,692 computed tomography volumes (CT-RATE dataset) \citep{hamamci_generalist_2026} in the latent space of a 3D variational autoencoder (MAISI) \citep{guo_maisi_2025}, which compresses full 3D CT volumes enough to make the double-backward pass tractable at useful spatial resolution, and evaluate the learned energy in two modes. The first is unconditional volumetric generation. The second evaluation uses the explicit scalar in regularizing an ill-posed inverse problem: sparse-view CBCT reconstruction from only 20 of 600 projections using Model-Based Iterative Reconstruction (MBIR) \citep{venkatakrishnan_plug-and-play_2013}, in which the same unconditionally trained potential enters a measurement-consistency loop as a 3D prior.

Our main contributions are summarized as follows:
\begin{itemize}
    \item \textbf{Projection improves the training signal:} We train a scalar potential by projecting the velocity field of a frozen, time-independent flow teacher onto the conservative subspace, replacing the stochastic transport targets of Energy Matching with deterministic supervision. A gradient-noise analysis on CIFAR-10 shows that the projected objective yields a stronger training signal than learning the potential from scratch, with improved convergence.
    \item \textbf{Energy-based priors at volumetric scale:} We train our PEM model on 25,692 CT-RATE volumes ($128\times128\times64$) in the latent space of the MAISI autoencoder, where a deterministic training objective makes the training feasible with our computation budget, and the resulting potential serves as a zero-shot prior for out-of-distribution detection and sparse-view cone-beam CT reconstruction.
\end{itemize}
\section{Methodology}
\label{sec:method}

Let $\mathbf{x} \in \mathbb{R}^D$ denote a state in the high-dimensional data space. We propose a framework for training an Energy Matching model on high-dimensional data by leveraging the transport efficiency of flow models. Our approach unfolds in three phases: (1) training a flow teacher to learn global transport; (2) projecting this teacher velocity field onto the conservative subspace of a scalar potential; and (3) refining the energy landscape using an Energy Matching contrastive loss. 

\subsection{Phase 1: Flow Teacher}
\label{subsec:phase1}
To provide a stable target for the projection, we first learn a transport map between a Gaussian prior $p_0$ and the data distribution $p_1$. We employ a transport objective, which learns a velocity field $v_{\text{teacher}}$ that creates straight paths between distributions following \citep{balcerak_energy_2025}.

To minimize the curvature of the transport trajectories and resolve path crossings, we utilize minibatch optimal transport (OT) couplings \citep{tong_improving_2023}. Instead of random pairings, we compute the discrete OT map between batches of the Gaussian prior $p_0$ and the data distribution $p_1$, storing these distance-minimizing pairs $(\mathbf{x}_0, \mathbf{x}_1)$ in a memory bank $\mathcal{M}$. We then train our flow teacher to match these straight conditional paths:
\begin{equation}
    \mathcal{L}_{\text{OT}} = \mathbb{E}_{t, (\mathbf{x}_0, \mathbf{x}_1) \sim \mathcal{M}} \left[ \| v_{\text{teacher}}(\mathbf{x}_t) - (\mathbf{x}_1 - \mathbf{x}_0) \|^2 \right],
\end{equation}
where $\mathbf{x}_t = t\mathbf{x}_1 + (1-t)\mathbf{x}_0$. By enforcing OT pairings, we drastically reduce trajectory intersections. This ensures the resulting marginal velocity field approaches a time-independent (autonomous) state ($\nabla_t v \approx 0$), providing a stable, low-variance target for the subsequent projection phase.

\subsection{Phase 2: Projection}
\label{subsec:phase2}

We now recover a scalar energy $\phi_\theta(\mathbf{x})$ whose gradient reproduces the teacher dynamics, $-\nabla_{\mathbf{x}} \phi_\theta \propto v_{\text{teacher}}$. The model is parametrized as a potential and differentiated with respect to its input, so the induced velocity field
\begin{equation}
    v_{\theta}(\mathbf{x}) = -N \nabla_{\mathbf{x}} \phi_\theta(\mathbf{x})
\end{equation}
is conservative by construction and $N$ represents the spatial dimension factor. The teacher, in contrast, was trained without this restriction. Minimizing the squared error between the two therefore does not fit the teacher pointwise but it returns the closest conservative field in $L^2$, weighted by the density of interpolants $\mathbf{x}_t$ along the transport paths. This is why we call the method \textit{projected}, and why the OT couplings of Phase 1 are crucial, as the more closely the teacher approximates straight paths, the more of its transport structure $\phi_\theta$ retains.

The projection objective is a mean squared error of the scalar potential against the frozen teacher: 
\begin{equation}     
\label{eq:proj_loss}
\mathcal{L}_{\text{proj}} = \mathbb{E}_{t, \mathbf{x}_t} \left[ \left\| -N \nabla_{\mathbf{x}} \phi_\theta(\mathbf{x}_t) - \text{sg}[v_{\text{teacher}}(\mathbf{x}_t)] \right\|^2 \right],
\end{equation} 
where $\text{sg}[\cdot]$ denotes the stop gradient operator. 

This phase is efficient despite the double-backward pass because its target is a deterministic function of $x_t$, not a single-sample estimate of $\mathbb{E}\left[ (x_1 - x_0) \;\middle|\; x_t \right]$. Both losses share a minimizer, but the projected loss removes the dominant variance term of learning transport from scratch. We quantify the resulting difference in gradient noise structure in Sec.~\ref{sec:experiments}, and it is what makes the total step count affordable for us at 3D scale. The complete procedure is summarized in the Appendix \ref{alg:projection} Algorithm \ref{alg:projection}.

\subsection{Phase 3: Energy Matching Refinement}
\label{subsec:phase3}

\begin{algorithm}[t]
\small
\caption{Projected Energy Matching with Teacher Supervision.}
\label{alg:energy_matching}
\begin{algorithmic}[1]
\Require Projected Energy Matching potential $\phi_\theta$ (from Phase 2), frozen teacher flow $v_\psi$ (from Phase 1), contrastive weight $\lambda_{\text{contrastive}}$, dimension factor $N = D$
\Require Gradient accumulation steps $K$, Langevin steps $M_{\text{Langevin}}$, step size $\Delta t$
\Require Standard Gaussian noise vectors $\eta \sim \mathcal{N}(0, I)$

\For{iteration $n = 0, 1, \ldots$}
    \State $g \leftarrow 0$ \Comment{Initialize gradient accumulator}
    \State $\mathcal{X}_{\text{cached}} \leftarrow \emptyset$ \Comment{Clear negative sample cache}

    \For{micro-batch $k = 1$ to $K$}
        \State Sample mini-batch $\{x_{\mathrm{data}}^{(j)}\}_{j=1}^B \sim \mathcal{D}$ and $\{x_{\mathrm{noise}}^{(j)}\}_{j=1}^B \sim \mathcal{N}(0, I)$
        \State Sample independent time steps $\{t^{(j)}\}_{j=1}^B \sim \mathcal{U}[0, 1]$
        \State $x_t^{(j)} \gets t^{(j)} x_{\mathrm{data}}^{(j)} + (1-t^{(j)}) x_{\mathrm{noise}}^{(j)}$ \Comment{Interpolate along optimal transport paths}

        \State $v_{\text{target}}^{(j)} \gets \text{sg}[v_{\psi}(x_t^{(j)})]$ \Comment{Frozen teacher velocity, single forward pass, no gradient}
        \State $v_{\text{base}}^{(j)} \gets -N \nabla_x \phi_\theta(x_t^{(j)})$ \Comment{Conservative flow}
        \State $\mathcal{L}_{\text{proj}} \gets \frac{1}{B} \sum_{j=1}^B \| v_{\text{base}}^{(j)} - v_{\text{target}}^{(j)} \|_2^2$

        \If{$\mathcal{X}_{\text{cached}} = \emptyset$} \Comment{Perform MCMC only on first micro-batch}
            \State Initialize negative samples $x_{\mathrm{neg}}^{(0)}$ from noise distribution
            \For{$m = 0, 1, \ldots, M_{\text{Langevin}}-1$}
                \State Sample independent noise $\eta^{(m)} \sim \mathcal{N}(0, I)$
                \State $x_{\mathrm{neg}}^{(m+1)} \gets x_{\mathrm{neg}}^{(m)} - \Delta t \times N \nabla_x \phi_{\theta}(x_{\mathrm{neg}}^{(m)}) + \sqrt{2 \Delta t}\,\eta^{(m)}$
            \EndFor
            \State $\mathcal{X}_{\text{cached}} \gets \text{sg}[x_{\mathrm{neg}}^{(M_{\text{Langevin}})}]$ \Comment{Detach and cache negative pool}
        \EndIf

        \State $x_{\mathrm{neg}} \gets \mathcal{X}_{\text{cached}}$ \Comment{Retrieve cached negatives}

        \State $\mathcal{L}_{\text{contrastive}} \gets \frac{1}{B} \sum_{j=1}^B \Big( \phi_{\theta}(x_{\mathrm{data}}^{(j)}) - \phi_{\theta}(x_{\mathrm{neg}}^{(j)}) \Big)$ \Comment{Raw potential values, no $N$}

        \State $\mathcal{L}_{\text{total}} \gets \mathcal{L}_{\text{proj}} + \lambda_{\text{contrastive}} \mathcal{L}_{\text{contrastive}}$
        \State $g \gets g + \frac{1}{K} \nabla_\theta \mathcal{L}_{\text{total}}$ \Comment{Accumulate gradients across micro-batches}
    \EndFor

    \State Update $\theta \gets \text{Optimizer}(\theta, g)$ \Comment{Apply synchronized gradient update}
\EndFor

\State \textbf{Return} Refined potential $\phi_\theta$
\end{algorithmic}
\end{algorithm}

The projection supplies the global optimal transport structure, but the resulting potential must be fine-tuned near the data manifold. In this final phase we refine $\phi_\theta$ using the Energy Matching framework \citep{balcerak_energy_2025}, retaining the frozen teacher as the transport supervisor. This joint objective combines the projection loss of Phase 2 ($\mathcal{L}_{\text{proj}}$), which maintains the transport, and a contrastive loss ($\mathcal{L}_{\text{contrastive}}$) driven by Langevin dynamics close to the data distribution. Keeping the teacher in the loop means the transport term remains a deterministic.

To adapt this framework for the massive memory footprint of high-dimensional 3D voxel grids, we introduce one efficiency modification to the standard Energy Matching pipeline. To manage VRAM limitations, 3D training requires gradient accumulation, but running deep Langevin chains for every micro-batch is computationally prohibitive. We solve this via \textit{Negative Caching}: we execute the expensive MCMC sampling only during the first micro-batch, cache these generated negative samples, and reuse them as a frozen negative pool for the remaining accumulation steps.

The two phases share a single objective, $\mathcal{L}_{\text{total}} = \mathcal{L}_{\text{proj}} + \lambda_{\text{contrastive}} \mathcal{L}_{\text{contrastive}}$.

\textbf{Transport ($\mathcal{L}_{\text{proj}}$):} We ensure the energy landscape retains the macro-level ``funnel'' shape required for efficient sampling from noise by continuing to regress the conservative field onto the frozen teacher as described in Eq. \ref{eq:proj_loss}. 

\textbf{Local Refinement ($\mathcal{L}_{\text{contrastive}}$):} Concurrently, near the data manifold, we transition to a diffusive regime, where we add scheduled noise, to capture high-frequency local modes. We carve out localized basins of attraction using a contrastive loss:

\begin{equation}
    \mathcal{L}_{\text{contrastive}} = \mathbb{E}_{\mathbf{x}^+ \sim \mathcal{D}} [\phi_\theta(\mathbf{x}^+)] - \mathbb{E}_{\mathbf{x}^- \sim p_\theta} [\phi_\theta(\mathbf{x}^-)]
\end{equation}
where $p_\theta(\mathbf{x}) \propto \exp(-\phi_\theta(\mathbf{x}))$ represents the model's currently learned distribution, and negatives $\mathbf{x}^-$ are drawn from it via Langevin dynamics on $\phi_\theta$.
\section{Experiments}
\label{sec:experiments}

We first examine whether PEM improves the training signal and convergence of energy models on CIFAR-10. We then scale the method to 3D CT-RATE volumes, where the resulting prior is evaluated for generation, detecting distribution shifts, and sparse-view CBCT reconstruction. These applications test whether the same learned distribution can support different inference tasks without retraining.

\subsection{Training Signal Quality \& Dynamics in 2D}
\label{subsec:cifar_setup}
\label{subsec:gradient_noise}
\label{subsec:cifar_fidelity}

Each PEM run is paired with an Energy Matching (EM) run trained from scratch under identical architecture, initialisation, data ordering, optimiser, schedule, and seed. The only difference is the regression target: frozen teacher velocities for PEM and stochastic interpolant targets for EM. The experiment covers batch sizes $B \in \{8, 32, 128, 512\}$ and multiple learning rates, with three seeds at selected settings, giving $20$ matched run pairs.

\begin{figure}[!t]
    \centering
    \includegraphics[width=0.85\textwidth]{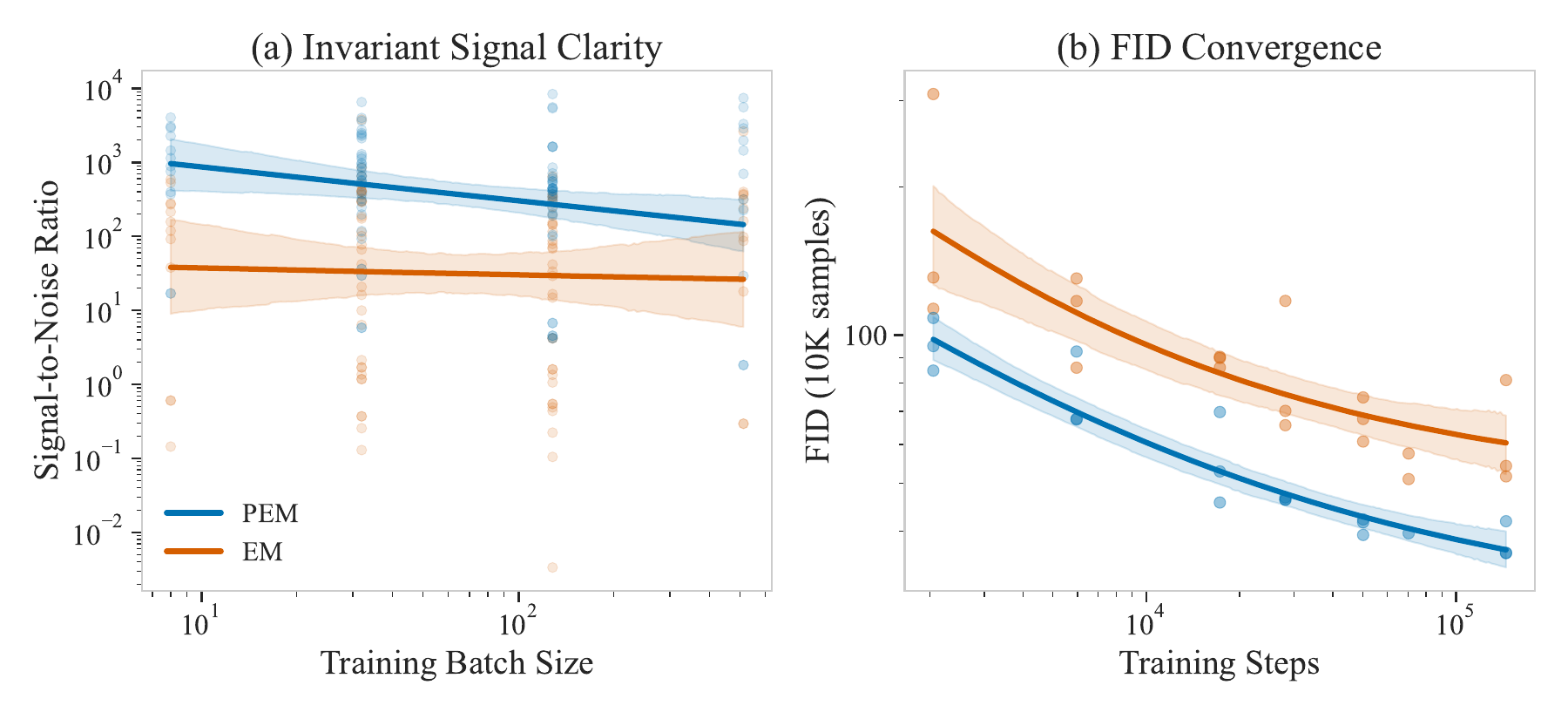}
    \caption{\textbf{Training signal and convergence on CIFAR-10.}
    \textbf{(a)} PEM exhibits higher gradient signal-to-noise ratios across the evaluated batch sizes.
    \textbf{(b)} FID during transport training shows that PEM converges faster than EM.
    Solid lines show posterior median fitted curves, and shaded regions indicate $95\%$ credible intervals for the fitted mean.}
    \label{fig:gradient_noise}
\end{figure}

Gradient-noise measurements \citep{mccandlish_empirical_2018} show a higher signal-to-noise ratio for PEM across the evaluated batch sizes (Fig.~\ref{fig:gradient_noise}a). The median squared norm of the mean parameter gradient is approximately $43\times$ larger than for EM. Gradient noise also increases, but by a smaller factor, so the average training signal becomes stronger relative to its variability.

This improvement is accompanied by better convergence behavior (Fig.~\ref{fig:gradient_noise}b). After processing approximately $450{,}000$ training examples per potential model, PEM achieves consistently lower FID as training progresses, suggesting faster convergence or better FID with the same compute budget. These results encourage us to learn a 3D generative prior in our next experiment under a constrained academic training budget, which otherwise seemed prohibitive. Statistical analysis details are provided in Appendix~\ref{app:bayes}.

\subsection{Evaluating the 3D Energy Prior}
\label{subsec:implementation_details}

\label{subsec:baseline_comparison}
\begin{table}[!t]
\centering
\small
\caption{\textbf{Unconditional 3D CT generation.}
FID evaluation uses $1{,}000$ held-out volumes of size $128\times128\times64$, with a sampling budget of $1{,}000$ function evaluations. The first three methods use the same U-Net backbone. The final row evaluates PEM with MedNeXt.}
\label{tab:main_results}
\begin{tabular}{l c c}
\toprule
\textbf{Method} & \textbf{Params} & \textbf{FID $\downarrow$} \\
\midrule
OT Flow Matching \citep{tong_improving_2023} & 10.3M & 42.94 \\
Rectified Flow \citep{liu_flow_2022} & 10.3M & 42.39 \\
PEM (Ours) & 10.7M & 40.87 \\
\midrule
PEM, MedNeXt backbone (Ours) & 10.2M & \textbf{35.87} \\
\bottomrule
\end{tabular}
\end{table}

To assess PEM at volumetric scale, we train an unconditional energy prior on CT-RATE volumes in the compressed latent space of the MAISI autoencoder. This representation makes training the 3D potential tractable. Dataset and preprocessing details are provided in Appendix~\ref{app:datasets_ctrate}.

\begin{figure}[H]
    \centering
    \includegraphics[width=0.92\textwidth]{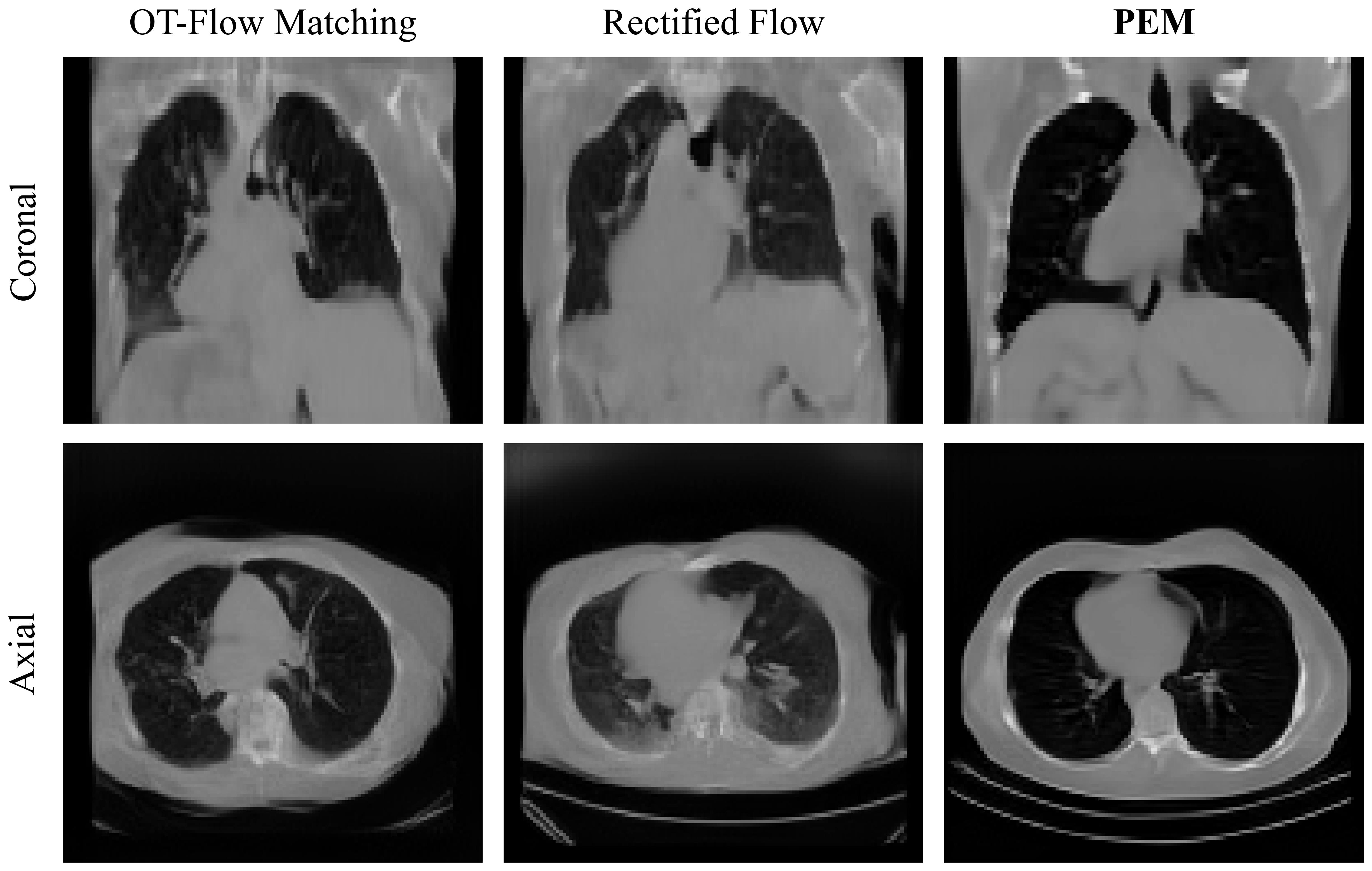}
    \caption{\textbf{Unconditionally generated CT volumes.}
    Coronal (top) and axial (bottom) slices from volumes generated by OT Flow Matching, Rectified Flow, and PEM using $1{,}000$ function evaluations.}
    \label{fig:slice_comparison}
\end{figure}

The resulting prior achieves generative quality competitive with time-dependent flow models (Table~\ref{tab:main_results}). Using the same U-Net backbone, PEM reaches an FID of $40.87$, compared with $42.94$ for OT Flow Matching and $42.39$ for Rectified Flow. Because PEM requires no time conditioning, the potential is compatible with backbones that lack timestep embeddings. With MedNeXt \citep{roy_mednext_2023}, PEM reaches an FID of $35.87$ with a comparable parameter count. FID is used to assess sample quality \citep{woodland_feature_2024}, and Figure~\ref{fig:slice_comparison} shows representative generated volumes. Ablations of the training phases are reported in Appendix~\ref{app:ablations}, Table~\ref{tab:ablation_results}.

\subsection{Detecting Distribution Shifts}

A learned prior can also help assess how supplied volumes differ from its training distribution. Here, the energy is used to distinguish familiar CT volumes from inputs affected by different acquisition conditions or reconstruction artifacts, without training a separate detector.

\begin{figure}[!t]
    \centering
    \includegraphics[width=\textwidth]{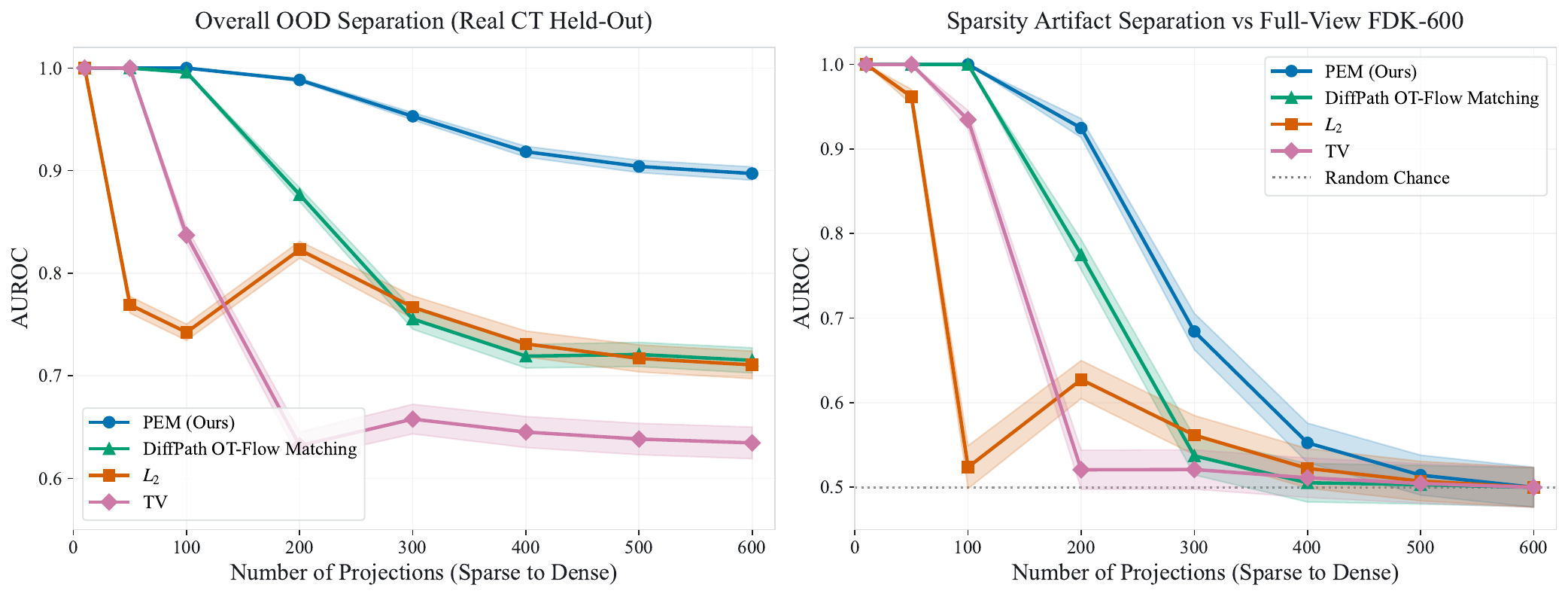}
\caption{\textbf{Scoring acquisition differences and reconstruction artifacts.}
We test whether the generative prior detects distribution shifts without additional training.
AUROC across projection counts for separating held-out CT-RATE volumes from CBCT inputs (left), and sparse-view FDK reconstructions from their $600$-projection FDK references (right).
PEM scores each volume with a single evaluation of the learned potential after encoding.
Higher values indicate better separation. At low projection counts, severe artifacts make separation straightforward for several scores.
Increasing projection counts reduces artifacts while keeping anatomy fixed, making the second task harder. PEM retains AUROC above $0.9$ at $200$ projections in this task.}
    \label{fig:ood_separation}
\end{figure}

Two settings test this capability across projection counts (Fig.~\ref{fig:ood_separation}). The first separates held-out CT-RATE volumes from real CBCT acquisitions. The second compares sparse-view reconstructions obtained with the Feldkamp-Davis-Kress (FDK) algorithm \citep{feldkamp_practical_1984} against their full-view, $600$-projection FDK references. In this second setting, reducing the number of projections introduces artifacts while keeping the underlying anatomy fixed.

Each volume is scored as $s(\mathbf{x})=\phi_\theta(\mathcal{E}(\mathbf{x}))$, where $\mathcal{E}$ is the MAISI encoder. After encoding, this requires a single evaluation of the potential. Comparisons use an OT Flow Matching trajectory score adapted from DiffPath \citep{heng_out--distribution_2024}, together with the $L_2$ norm and total variation (TV) of the volume.

At the lowest view counts, severe artifacts make separation straightforward for several scores. As additional projections improve reconstruction quality, the differences become less obvious. PEM retains stronger separation in this regime: in the artifact-detection task, its AUROC remains above $0.9$ at $200$ of $600$ projections, where the alternative scores are already less discriminative.

\begin{figure}[!t]
    \centering
    \includegraphics[width=0.8\textwidth]{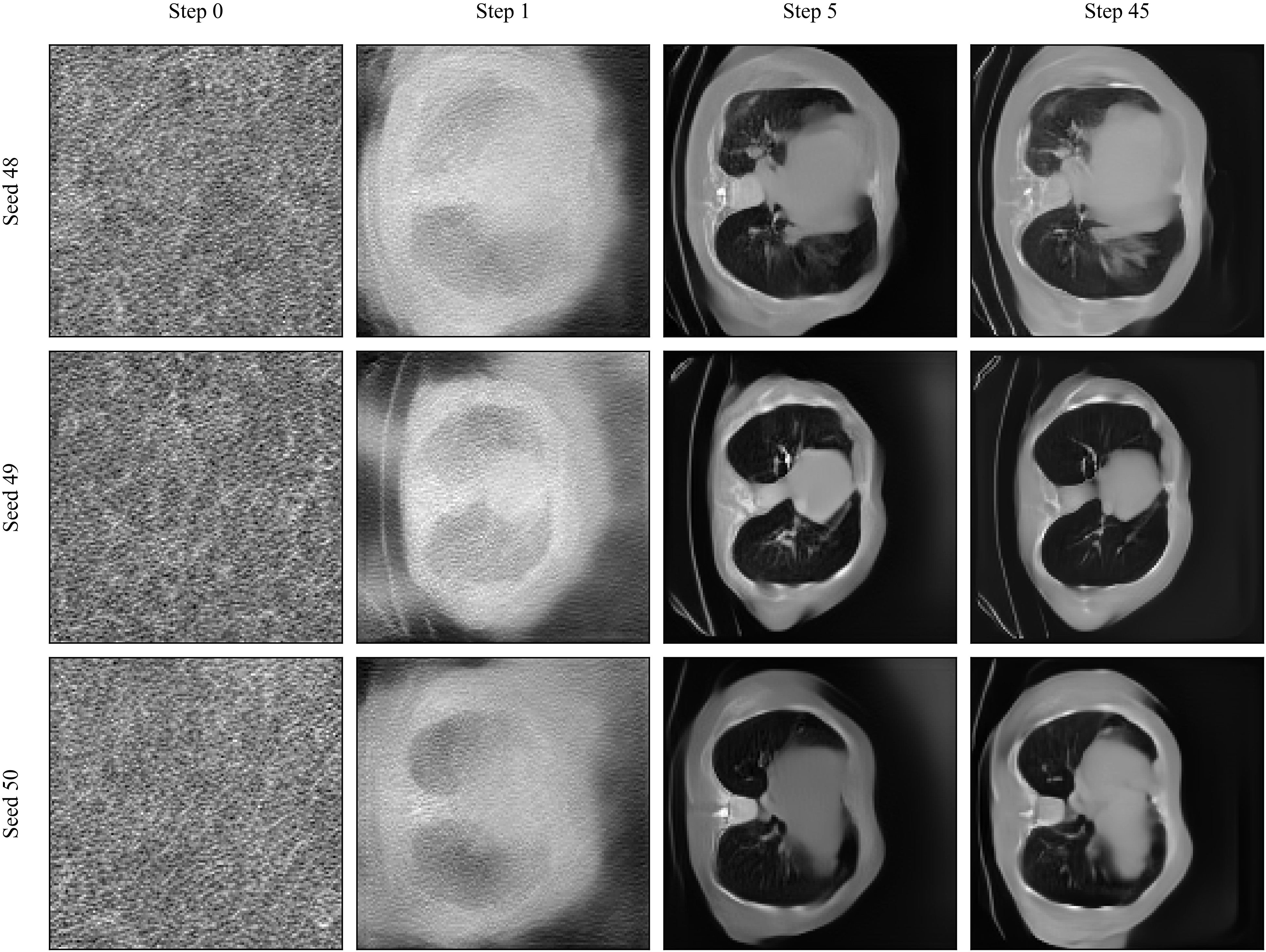}
    \caption{\textbf{Extended Manifold Exploration via Explicit Energy.} Unlike standard flow models that integrate forward from noise ($t=0$) and terminate rigidly at the data manifold ($t=1$), our PEM permits unbounded Langevin sampling. Starting from pure noise (Langevin Step 0), the global energy funnel rapidly projects the sample toward valid medical anatomy (Step 5). Continuing to sample extensively (Step 45) largely avoids the severe artifact accumulation typically seen when forcing time-dependent models past their endpoint.}
    \label{fig:manifold_exploration}
\end{figure}

\begin{table}[!t]
\centering
\small
\caption{\textbf{Sparse-View CBCT Reconstruction (20 of 600 projections).} Mean $\pm$ std over 10 held-out volumes. All iterative solvers share the forward operator, initialisation and step budget; only the regulariser differs. The energy prior admits a single conservative objective (Eq.~\ref{eq:map}) we show MAP estimation here.}
\label{tab:cbct_recon}
\begin{tabular}{lcccc}
\toprule
\textbf{Method} & \textbf{Prior} & \textbf{PSNR $\uparrow$} & \textbf{SSIM $\uparrow$} & \textbf{LPIPS $\downarrow$} \\
\midrule
FBP                                & --     & 20.91 $\pm$ 0.67          & 0.4455 $\pm$ 0.0122          & 1.0099 $\pm$ 0.0587 \\
Iterative TV                       & --     & 26.75 $\pm$ 0.87          & 0.7410 $\pm$ 0.0380          & 0.3375 $\pm$ 0.0369 \\
OT-Flow Matching MAP               & Flow   & 31.92 $\pm$ 1.22          & 0.8881 $\pm$ 0.0303          & 0.0960 $\pm$ 0.0244 \\
\textbf{Ours}, PEM variational MAP & Energy & \textbf{32.55 $\pm$ 1.30} & \textbf{0.9161 $\pm$ 0.0252} & \textbf{0.0621 $\pm$ 0.0154} \\
\bottomrule
\end{tabular}
\end{table}

\begin{figure}[!b]
    \centering
    \includegraphics[width=0.87\textwidth]{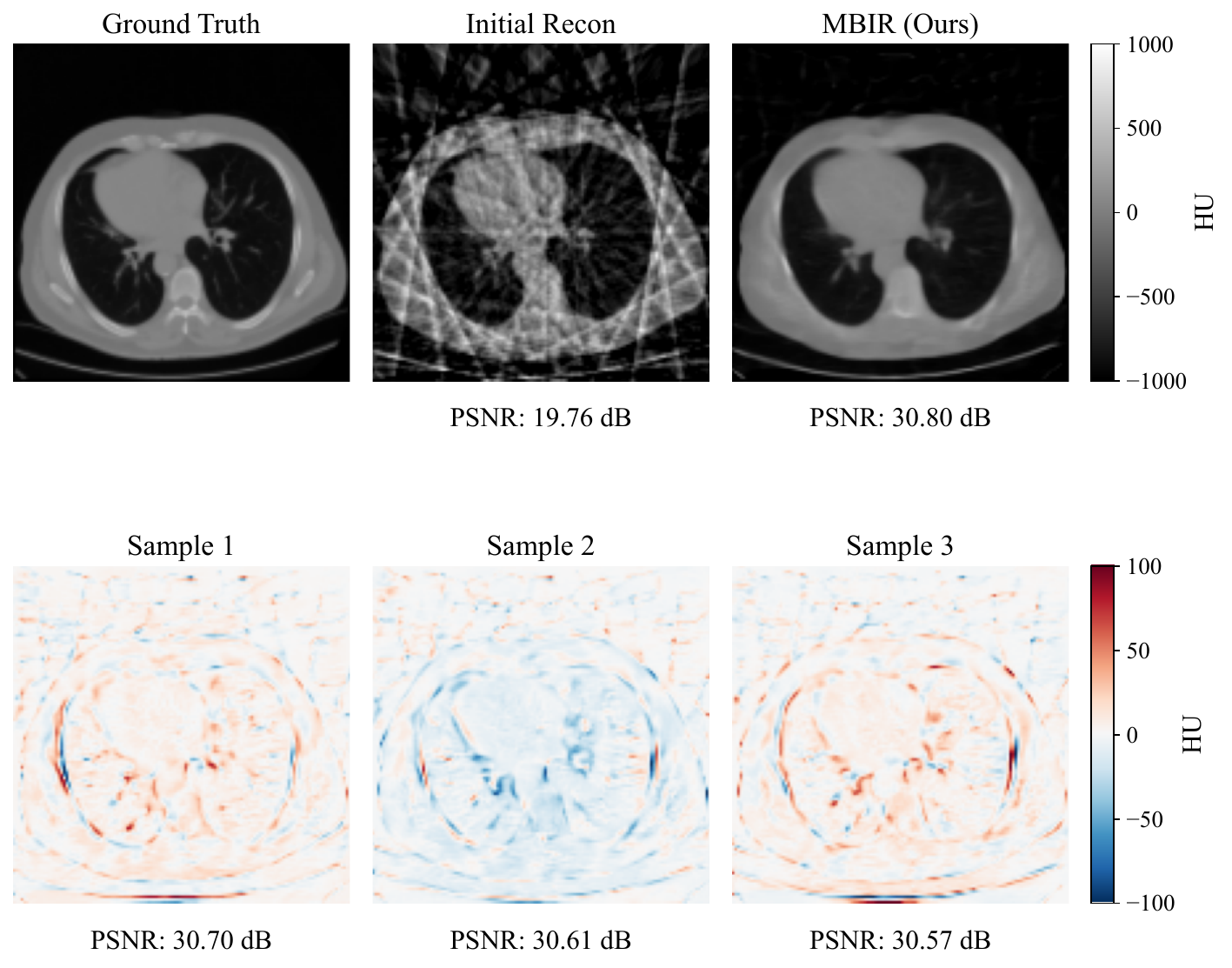}
    \caption{\textbf{Diversity under sparse measurements.}
    Top: ground truth, sparse-view initialisation, and ensemble-mean reconstruction.
    Bottom: three reconstructions obtained through Langevin sampling with approximate null-space repulsion, shown as differences from their ensemble mean. Bottom-row colours indicate intensity differences in Hounsfield units, and PSNR values refer to the underlying individual reconstructions.}
    \label{fig:diversity}
\end{figure}

\subsection{Manifold Exploration}
\label{subsec:manifold_exploration}

The learned prior supports exploration both with and without measurements. In the unconstrained case, Langevin dynamics can continue to evolve a generated volume under the same time-independent potential. Fig.~\ref{fig:manifold_exploration} illustrates the resulting anatomical variations and reports sample quality over extended sampling horizons (Appendix~\ref{app:gen_quality} Fig.~\ref{fig:extended_horizon_fid}).

Adding measurements turns this exploration into a reconstruction task. Given measurements $\mathbf{y}$ and a forward operator $\mathcal{A}$, we combine measurement consistency with the learned prior:
\begin{equation}
\label{eq:map}
    \mathbf{z}^\star = \arg\min_{\mathbf{z}} \;
    \| \mathcal{A}\,\mathcal{D}(\mathbf{z}) - \mathbf{y} \|^2
    + \lambda\,\phi_\theta(\mathbf{z}),
\end{equation}
where $\mathcal{D}$ is the MAISI decoder and $\lambda$ weights the prior. For sparse-view CBCT, we retain only $20$ of $600$ projections and optimise Eq.~\ref{eq:map} in latent space for $2500$ steps, without retraining the prior or introducing measurement-specific conditioning. For the MAP estimate, we minimise Eq.~\ref{eq:map} with Adam for 2500 steps, initialised from the encoded FBP reconstruction. For posterior sampling, we integrate Langevin dynamics on the same objective (Appendix \ref{app:mbir_details} \ref{eq:mbir_langevin}). Reconstruction results are shown in Appendix~\ref{app:gen_quality} Fig.~\ref{fig:cbct_reconstruction} and Table~\ref{tab:cbct_recon}, with solver details in Appendix~\ref{app:mbir_details}.

Sparse measurements still leave room for multiple reconstructions. To explore this ambiguity, we draw three posterior samples jointly with Langevin dynamics (Appendix \ref{app:mbir_details} Eq.~\ref{eq:mbir_langevin} and add a repulsion term \ref{app:repulsion}. We apply this repulsion in the approximate null space, encouraging variation in directions weakly constrained by the measurements (see Appendix~\ref{app:repulsion} for technical details). The bottom row of Figure~\ref{fig:diversity} shows the resulting differences from the ensemble mean. Across $10$ volumes and $3$ seeds, repulsion improves the ensemble-mean PSNR from $30.02$ to $30.22$\,dB compared with the same procedure without repulsion. Table~\ref{tab:cbct_diverse} reports the corresponding individual and ensemble reconstruction metrics.

\section{Discussion and Conclusion}
\label{sec:conclusion}

We introduced Projected Energy Matching (PEM) as a more stable route to learning time-independent energy-based models through transport-field supervision, as in Energy Matching. A frozen velocity teacher replaces stochastic transport targets with deterministic supervision, while contrastive refinement shapes the potential near the data manifold. On CIFAR-10, this approach yields a stronger training signal, improved seed stability, and lower FID in a compute-matched comparison excluding teacher pretraining.

By making the energy landscape easier to learn, PEM enables Energy Matching at volumetric scale. Our pipeline learns a 3D latent energy prior that achieves lower FID than flow baselines using the same backbone, particularly with an architecture that natively omits time embeddings.

\textbf{Limitations.} PEM substantially amortises the transport regime, while the computational cost of contrastive Langevin sampling and the memory demands of double-backward differentiation remain. Both increase with dimensionality, but are manageable in the volumetric latent space studied here.

\textbf{Broader Impacts.} PEM opens opportunities for wider use of energy-based models in high-dimensional scientific and medical applications. Their scalar formulation allows learned priors to be combined directly with physical constraints and measurement objectives, creating opportunities for controlled generation and inference without task-specific retraining. In CT imaging, sparse-view reconstruction offers a route toward fewer projections and lower radiation dose. Translating this potential into clinical benefit requires validating anatomical fidelity, particularly for structures weakly constrained by the measurements, and robustness beyond the CT-RATE training distribution.

A single unconditionally trained potential generates volumes, scores out-of-distribution inputs, and supports sparse-view CBCT. Its energy can be evaluated directly and combined with measurement objectives and physical constraints without retraining. By combining this flexibility with lower FID than flow baselines, PEM advances a broader paradigm for volumetric generative modeling in which explicit energy landscapes serve as a common foundation for generation and scientific inference.
 \section*{AI Use Statement} 
 ChatGPT and Claude were used to assist with drafting and polishing the manuscript text, and Claude Code assisted with method implementation, coding and debugging. The research ideas and derivations were developed by the authors. The authors reviewed the code and evaluation, checked the final wording and citations, and take responsibility for the correctness and originality of the final submission.

% --- BIBLIOGRAPHY ---
\bibliographystyle{plainnat}
\bibliography{references}

%%%%%%%%%%%%%%%%%%%%%%%%%%%%%%%%%%%%%%%%%%%%%%%%%%%%%%%%%%%%
% --- APPENDIX ---
\newpage
\appendix
\section{Appendix: Network Architectures and Implementation Details} 
\label{app:architecture}  
All models operate on the compressed 4-channel latent representation of the 3D volumes. We instantiate two backbones. The first, MedNeXt, is used for our headline results and balances the long-range spatial dependencies typically handled by Transformers with the inductive biases of CNNs. The second, a 3D diffusion U-Net, compares against competing methods under a backbone standard in the flow- and diffusion-matching literature, so reported differences are attributable to the objective rather than the architecture. Both backbones are used for the Phase-1 flow teacher and for the Phase-2/3 potential model, and both share an identical potential head.  

\subsection{Backbone A: MedNeXt} 
MedNeXt is a fully convolutional 3D architecture inspired by the ConvNeXt design, which scales CNNs using Transformer-driven principles \citep{roy_mednext_2023}. The core computational unit is the residual inverted bottleneck block. Each block consists of three layers: (1) a depthwise convolution to aggregate spatial context, (2) an expansion layer (with an expansion ratio of $R=2$) using $1 \times 1 \times 1$ convolutions and GELU activations to decouple width scaling from receptive field scaling, and (3) a compression layer projecting back to the original channel dimension \citep{roy_mednext_2023}. This backbone is time-independent: no timestep embedding is injected.  

\subsection{Backbone B: Diffusion U-Net (baseline-matched)} 
The second backbone is the standard 3D DiffusionModelUNet of MONAI \cite{cardoso_monai_2022}, with 4 input and output channels and 4 resolution levels of width $(16, 32, 64, 128)$. Each level contains 2 residual blocks with group normalisation over 16 groups. Self-attention is applied only at the lowest-resolution level, with 64 channels per attention head and attention logits upcast to \texttt{float32} for numerical stability; no cross-attention conditioning is used. Unlike Backbone A, this network receives a sinusoidal time step embedding which is only used for OT-Flow Matching \cite{lipman_flow_2022} and Rectified Flow \cite{liu_flow_2022}. For PEM trained with this backbone, the models are time-independent, which is achieved by passing a dummy $t=0.5$ on every pass.

\subsection{Phase 1: Unconstrained Flow Teacher} 
In the first phase (learning the Optimal Transport flow), the network must output an unconstrained 3D marginal vector field $v_{\text{teacher}}$.  For Backbone A we append a shallow \emph{Velocity Head}: a single 3D MedNeXt block (expansion ratio 2, $3\times3\times3$ kernel, group normalisation) followed by a $1\times1\times1$ convolution mapping back to the 4-channel latent dimension. For Backbone B the U-Net already emits a 4-channel tensor at full latent resolution, so the velocity head is the identity. In both cases the teacher is a highly expressive, unconstrained functional mapping $\mathbf{x}_t \mapsto v(\mathbf{x}_t)$.  

\subsection{Phases 2 and 3: Projection and Potential Training} 
During the projection phase, and in the subsequent potential refinement phase, the architecture must shift from a dense vector field to a single global scalar potential $\phi_\theta(\mathbf{x}_t)$. We replace the velocity head with a \emph{Potential Head} that is identical for both backbones, so that the two arms differ only in their feature extractor. The head is a small Transformer: 
\begin{enumerate}
    \item \textbf{Patch embedding:} a non-overlapping 3D patch embedding (patch size $2^3$) maps the 4-channel backbone output to a sequence of tokens of dimension $d=128$.
    \item \textbf{Global mixing:} two pre-norm Transformer encoder layers ($d=128$, 4 attention heads, feed-forward dimension 512, SiLU activation, no dropout), implemented with a \texttt{torch.compile}- and FlashAttention-compatible layer so that the double-backward pass through the head remains fused.
    \item \textbf{Scalar projection:} mean pooling over the token dimension followed by a linear map $\mathbb{R}^{128}\to\mathbb{R}$, yielding one scalar per sample, $\phi_\theta(\mathbf{x}_t) \in \mathbb{R}^{B}$. 
\end{enumerate} Self-attention is used here rather than further convolution because the potential is a single global functional of the volume: the head must aggregate evidence across the entire latent field, not refine local structure.  The forward pass of the functional velocity model computes the scalar $\phi_\theta(\mathbf{x}_t)$ via the potential model and then differentiates with respect to the input to output the conservative field $-\nabla_{\mathbf{x}}\phi_\theta(\mathbf{x}_t)$. Because the head is a scalar, the gradient is obtained in a single reverse-mode pass, and the training objective requires one further backward pass through that graph.
\section{Appendix: Datasets and Preprocessing Details}
\label{app:datasets}

To rigorously evaluate the scalability and physical accuracy of our Projected Energy Matching Model, we use CIFAR-10 as a low-dimensional, fully controlled testbed for the gradient-noise analysis, alongside a 3D CT-dataset that exhibit complex, continuous topologies and strict domain constraints. This section details the preprocessing pipelines, autoencoder compression.

\subsection{CIFAR-10: Low-Dimensional Variance Study}
\label{app:datasets_cifar10}

CIFAR-10 \citep{krizhevsky_learning_2009} consists of 60,000 RGB images at $32 \times 32$ resolution, split into 50,000 training and 10,000 test images over 10 classes. We use it exclusively for the gradient-noise and critical-batch-size analysis (Section~\ref{subsec:gradient_noise}), where the low dimensionality makes the paired scratch/projected sweep over batch sizes, learning rates, and seeds tractable. Class labels are discarded; all models are trained unconditionally. The 50,000 training images are further split 90/10 into 45,000 training and 5,000 held-out images, the latter providing the shared evaluation split used for the flow-matching loss closure reported across arms.

\paragraph{Preprocessing.}
Images are used at their native $32 \times 32 \times 3$ resolution, so both the teacher velocity field and the student potential operate directly in pixel space. The pipeline consists of random horizontal flipping, followed by an intensity rescaling from $[0, 255]$ to $[-1, 1]$ via per-channel normalization with mean and standard deviation $0.5$. Samples are mapped back to $[0, 1]$ at generation time for evaluation.

\subsection{CT-RATE: High-Dimensional 3D Medical Volumes}
\label{app:datasets_ctrate}

The CT-RATE dataset \citep{hamamci_generalist_2026} is a large-scale, real-world cohort of volumetric chest CT scans, providing complex, continuous 3D anatomies. The dataset consists of 25,692 non-contrast chest CT volumes (expanded to 50,188 through various reconstructions) originating from 21,304 unique patients. Following \cite{hamamci_generalist_2026}, we divided the cohort at the patient level: 20,000 patients were allocated to the training set, and the remaining 1,304 patients were reserved for the validation set.

\paragraph{Preprocessing and Normalization.}
To standardize the volumes for neural network processing, we employ a rigorous spatial and intensity normalization pipeline. First, all scans are spatially resampled to a voxel spacing of $1.5 \times 1.5 \times 2.367 \text{ mm}^3$ using trilinear interpolation. To achieve a uniform high-resolution grid of $128 \times 128 \times 64$ voxels, we center-crop or pad the axial (XY) plane. If the z-axis (depth) exceeds 64 slices, it is center-cropped; if it is shorter, it is padded by replicating the first and last boundary slices. 

\paragraph{Latent Space Embedding via MAISI.}
We compress the preprocessed volumes using a 3D Variational Auto Encoder (VAE) into a latent space \cite{guo_maisi_2025}. 

The model is trained on a combination of objectives to ensure the volume reconstructions adhere closely to the image manifold with high local realism. The overall objective integrates a voxel-space $L_1$ reconstruction loss ($L_{\text{recon}}$), an LPIPS perceptual loss ($L_{\text{lpips}}$), an adversarial loss ($L_{\text{adv}}$) utilizing a 3D discriminator to penalize unrealistic artifacts, and a Kullback-Leibler (KL) regularization term ($L_{\text{reg}}$) to enforce a standard normal distribution on the latent features, preventing high-variance latent spaces.

\section{Appendix: Generative Quality}
\label{app:gen_quality}

\subsection{Ill-Posed Inverse Problem: Sparse-View CBCT Reconstruction}
\label{app_subsec:illposed_cbct}
\begin{figure}[t!]
    \centering
    \includegraphics[width=0.87\textwidth]{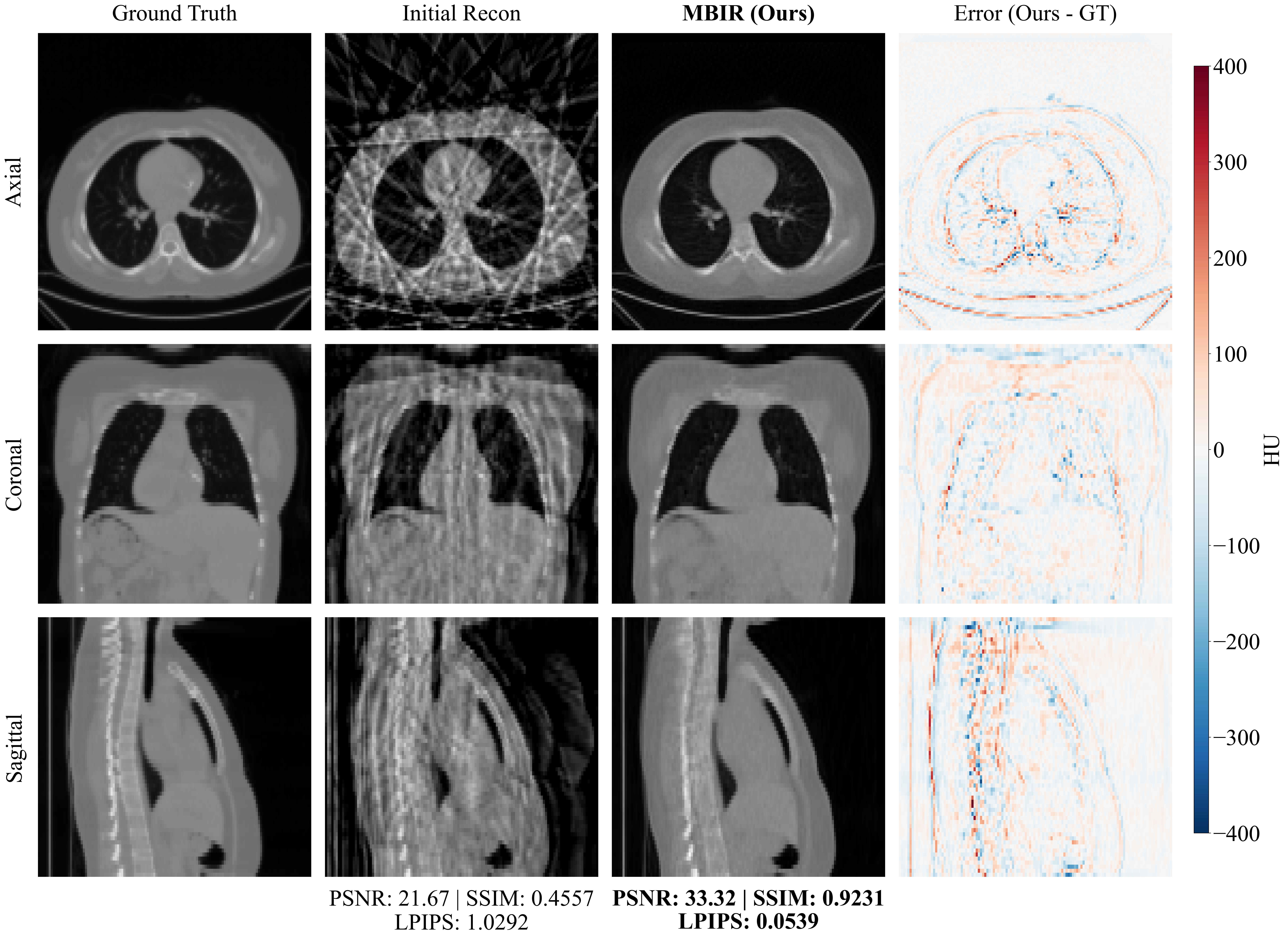}
    \caption{\textbf{Sparse-View CBCT Reconstruction via MBIR.} Qualitative evaluation on an extreme ill-posed inverse problem (20 of 600 projections).\textbf{Left:} full-view ground truth. \textbf{Middle:} the initial analytical reconstruction, heavily corrupted by sparse-view streak artifacts. \textbf{Right:} a \emph{MAP sample} obtained by a joint objective of Eq.~\ref{eq:map} ($33.32$\,dB, SSIM $0.9231$, LPIPS $0.0539$)}
    \label{fig:cbct_reconstruction}
\end{figure}

We evaluate the trained PEM as a zero-shot prior for sparse-view cone-beam CT (CBCT) reconstruction, a severely ill-posed inverse problem, using only 20 of 600 projections (Fig. \ref{fig:cbct_reconstruction}).

\subsection{Manifold Exploration}
\label{app_subsec:manifold_explore}

\begin{figure}[t!]
    \centering
    \includegraphics[width=0.8\textwidth]{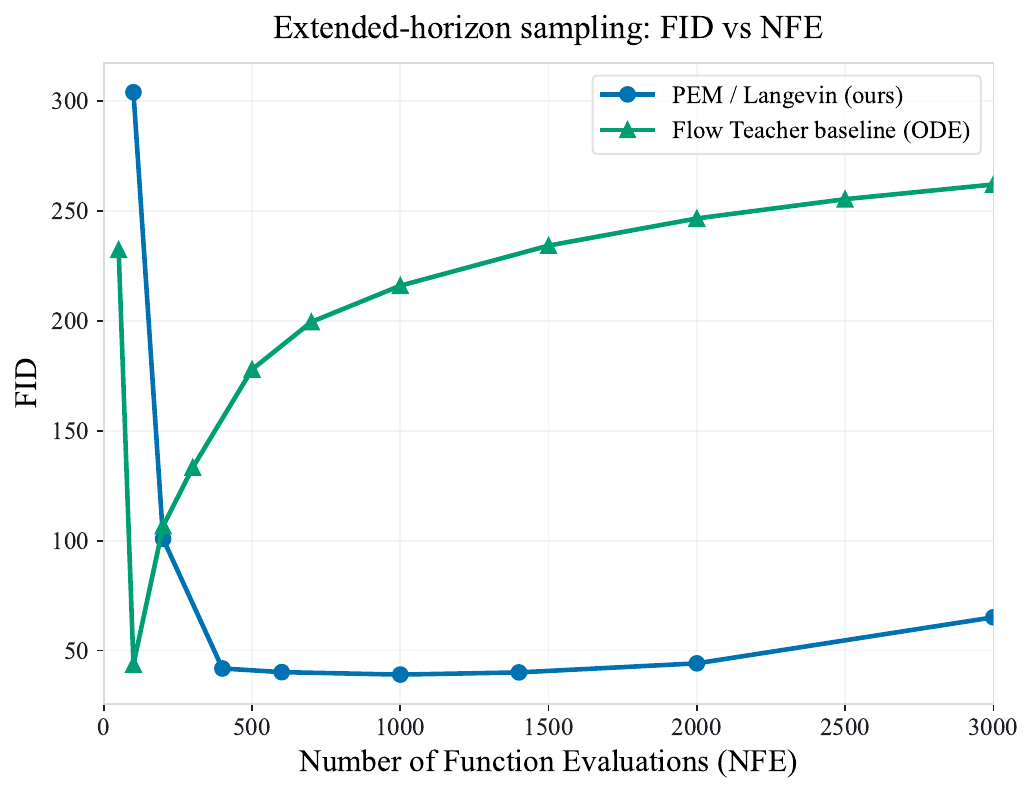}
    \caption{\textbf{Extended-Horizon Sample Quality.} Quantitative evaluation of Fréchet Inception Distance (FID) across cumulative evaluation costs (NFE). OT Flow Matching model (green) degrades sharply when the deterministic ODEs are integrated past its native $t=1$ horizon. In contrast, our Projected Energy Matching (PEM, blue) safely permits unbounded Langevin sampling. By continuing to sample well beyond standard limits (extending past 2000 NFE), PEM maintains a stable, low FID, successfully avoiding the severe artifact accumulation typically seen when forcing time-dependent models past their endpoint.}
    \label{fig:extended_horizon_fid}
\end{figure}

Since PEM is time-independent, Langevin sampling on the learned 3D scalar potential can continue far beyond the flow endpoint while staying on the data manifold (Fig.~\ref{fig:extended_horizon_fid}).

\section{Appendix: Algorithms}
\label{app:algo}

\begin{algorithm}[H]
\small
\caption{Projection of the Flow Teacher (Phase 2).}
\label{alg:projection}
\begin{algorithmic}[1]
\Require Pre-trained teacher flow $v_{\psi}$ (frozen)
\Require Potential $\phi_\theta$, spatial dimension factor $N = D$

\State Initialize $\theta$ randomly

\For{iteration $0, 1, \ldots$}
    \State Sample mini-batch $\{x_{\mathrm{data}}^{(j)}\}_{j=1}^B \sim \mathcal{D}$ and $\{x_{\mathrm{noise}}^{(j)}\}_{j=1}^B \sim \mathcal{N}(0, I)$
    \State Sample independent time steps $\{t^{(j)}\}_{j=1}^B \sim \mathcal{U}[0, 1]$
    \State $x_t^{(j)} \gets t^{(j)} x_{\mathrm{data}}^{(j)} + (1-t^{(j)}) x_{\mathrm{noise}}^{(j)}$ \Comment{Interpolate along optimal transport paths}

    \State $v_{\text{teacher}} \gets v_{\psi}(x_t)$ \Comment{Single forward pass, no gradient}

    \State $v_{\text{base}} \gets -N \nabla_{x} \phi_\theta(x_t)$ \Comment{Conservative field by construction}
    \State $\mathcal{L}_{\text{proj}} \gets \frac{1}{B} \sum_{j=1}^B \| v_{\text{base}}^{(j)} - \text{sg}[v_{\text{teacher}}^{(j)}] \|_2^2$
    \State Update $\theta \gets \text{Optimizer}(\nabla_\theta \mathcal{L}_{\text{proj}})$ \Comment{Double-backward through $\nabla_x \phi_\theta$}
\EndFor

\State \textbf{Return} conservative potential $\phi_\theta$
\end{algorithmic}
\end{algorithm}
\section{Appendix: Ablations}
\label{app:ablations}

\begin{table}[H]
\centering
\small
\caption{Quantitative ablation of the progressive UNet training phases on the CT-Rate dataset. The table reports parameter counts and the resulting generative fidelity at each phase of the pipeline.}
\label{tab:ablation_results}
\begin{tabular}{l c c}
\toprule
\textbf{Method} & \textbf{Params} & \textbf{FID $\downarrow$} \\
\midrule
PEM Phase 1 & 10.3M & 41.90 \\
PEM Phases 1+2 & 10.7M & 62.60 \\
PEM Phases 1+2+3 & 10.7M & 40.87 \\
\bottomrule
\end{tabular}
\end{table}

\section{Bayesian Analysis Details}\label{app:bayes}

All intervals are 95\% highest-density intervals. Paired effects (Section \ref{subsec:cifar_setup} - \ref{subsec:cifar_fidelity}) use conjugate or closed-form posteriors; the curves in Fig. \ref{fig:gradient_noise} are fitted with NUTS in PyMC \cite{abril-pla_pymc_2023}.

\textbf{Critical batch size.} For each checkpoint $j$ of run $r$ admissible in both arms ($n = \textbf{72}$ pairs, $\textbf{8}$ runs), let $\delta_{rj}$ be the difference in $\log_{10} B_{\mathrm{crit}}$ between PEM and EM. Checkpoints within a run are dependent, so we pool them through a run-level effect:
\begin{equation}
\delta_{rj} \sim \mathcal{N}(\theta_r, \sigma^2), \quad \theta_r \sim \mathcal{N}(\mu, \tau^2), \quad \mu \sim \mathcal{N}(0, 10^2), \quad \sigma^2, \tau^2 \sim \mathrm{InvGamma}(2, 1),
\end{equation}
and report $10^{\mu}$. We sample with Gibbs (40{,}000 iterations, 10{,}000 burn-in).

\textbf{FID.} Final-checkpoint log FID differences use a normal model with Jeffreys prior, whose posterior for the mean is Student-$t$. For the $B = 32$ long runs, the same hierarchical model as above is fitted with seed as the group. Seed dispersion at $B \in \{32, 128\}$ uses the scaled inverse-$\chi^2$ posterior of the residual variance ($\nu = 4$); we report the ratio of standard deviations.

\textbf{Admissibility.} Beta-Binomial per arm with a $\mathrm{Beta}(1,1)$ prior.

\textbf{Figure \ref{fig:gradient_noise}} (a) Per arm, $\log_{10}$ SNR is regressed on $\log B$ and $\log t$ with run intercepts and a Student-$t$ likelihood. (b) Per arm, FID follows a saturating power law $F_r + A_r (t/t_0)^{-\alpha_r}$ with run-level parameters, fitted on $\log_{10}$ FID with a Student-$t$ likelihood. Lines are posterior medians averaged over observed runs; bands are 95\% HDIs. We use 4 chains of 2{,}000 draws after 3{,}000 tuning steps, with $\hat R \le 1.01$ and no divergences.
\section{Appendix: Model-Based Iterative Reconstruction (MBIR) Frameworks}
\label{app:mbir_details}

In this work, we deploy our learned 3D priors to solve ill-posed medical inverse problems, specifically Model-Based Iterative Reconstruction (MBIR) for sparse-view CBCT in latent space. 
\subsection{Forward Model and Data Consistency}
The inference approach utilizes a physical forward modeling pipeline built on CTorch \cite{jiang_ctorch_2025}. The data consistency loss $\mathcal{L}_{\text{data}}(\mathbf{z})$ is formulated depending on the clinical scenario:
\begin{itemize}
    \item \textbf{Sparse-View CT:} We utilize a standard Mean Squared Error (MSE) loss against the measured sparse sinogram $\mathbf{y}_{\text{meas}}$: $\mathcal{L}_{\text{data}} = \frac{1}{2} \| \mathbf{y}_{\text{meas}} - A(\mathcal{D}(\mathbf{z})) \|_2^2$.
\end{itemize}

\subsection{Posterior Sampling via Explicit Langevin Dynamics}
Our primary method leverages the explicit nature of our Projected Energy Matching to perform unconstrained posterior sampling. Instead of directly optimizing for a single point estimate, we sample from the posterior distribution $p(\mathbf{z} | \mathbf{y}) \propto p(\mathbf{y} | \mathbf{z}) p(\mathbf{z})$ using Langevin dynamics.

As implemented in our solver, we initialize the latent state with random noise and integrate using a Heun-based Stochastic Differential Equation (SDE) solver \cite{balcerak_energy_2025}. The update step is guided by the joint gradients of the explicit energy prior and the physical data loss:
\begin{equation}
\label{eq:mbir_langevin}
    \mathbf{z}_{k+1} = \mathbf{z}_k - \gamma \nabla_{\mathbf{z}} \left( \lambda_{\text{data}} \mathcal{L}_{\text{data}}(\mathbf{z}_k) + \phi_\theta(\mathbf{z}_k) \right) + \sqrt{2\gamma}\boldsymbol{\epsilon}
\end{equation}
where $\gamma$ is the step size, $\lambda_{\text{data}}$ controls the relative strength of the physical constraints, and $\phi_\theta(\mathbf{z})$ is the learned explicit scalar potential. The sampling procedure allows the state to settle into a low-energy basin on the data manifold while strictly adhering to the projection geometries.

\section{Appendix: Repulsion}
\label{app:repulsion}
We optimise $K=3$ latents $\mathbf{z}_1,\dots,\mathbf{z}_K$ jointly, initialised from the same FBP reconstruction plus Gaussian noise ($\sigma=0.01$), with decoded volumes $\mathbf{x}_k=\mathcal{D}(\mathbf{z}_k)$. Each $\mathbf{z}_k$ descends Eq.~\ref{eq:map} plus the repulsion term following \cite{balcerak_energy_2025}
\begin{equation}
\label{eq:repulsion}
\mathcal{L}_{\mathrm{rep}}=\sum_{i<j}\exp\!\left(-\frac{\|\mathcal{B}(\mathbf{x}_i-\mathbf{x}_j)\|_2^2}{2h^2}\right),\qquad \mathcal{B}=\mathcal{I}-\mathcal{R}\mathcal{A},
\end{equation}
where $\mathcal{R}$ is filtered back-projection. $\mathcal{B}$ removes the part of a volume that is recoverable from its projections, so $\|\mathcal{B}(\mathbf{x}_i-\mathbf{x}_j)\|_2$ is the distance between two particles in the (approximate) null space of $\mathcal{A}$, and $\mathcal{L}_{\mathrm{rep}}$ separates them only where the measurements are uninformative. The bandwidth $h$ is the median of these pairwise distances, recomputed during optimisation and not differentiated. The gradient $\nabla_{\mathbf{z}_k}\mathcal{L}_{\mathrm{rep}}$ is rescaled to norm $w_{\mathrm{rep}}\|\nabla_{\mathbf{z}_k}\phi_\theta\|_2$ with $w_{\mathrm{rep}}=0.1$, so repulsion cannot override the prior. The baseline is identical with $w_{\mathrm{rep}}=0$, i.e.\ particles differ only through their initial noise. We report metrics of single particles, of the ensemble mean $\bar{\mathbf{x}}=\frac{1}{K}\sum_k\mathbf{x}_k$, and of the best particle selected with the ground truth (best-of-$K$). Calibration $r$ is the Pearson correlation over voxels between the ensemble standard deviation and $|\bar{\mathbf{x}}-\mathbf{x}_{\mathrm{gt}}|$.

\begin{table}[H]
\centering
\caption{Sparse-view CBCT ($20/600$ views), $K=3$ particles sampled with Langevin posterior sampling with and without null-space repulsion. Mean over $10$ volumes $\times$ $3$ seeds.}
\label{tab:cbct_diverse}
\begin{tabular}{lccc}
\toprule
 & PSNR $\uparrow$ & SSIM $\uparrow$ & LPIPS $\downarrow$ \\
\midrule
\multicolumn{4}{l}{\textit{Without repulsion}} \\
Single particle & $29.88$ & $0.808$ & $0.202$ \\
Ensemble mean & $30.02$ & $0.815$ & $0.198$ \\
\midrule
\multicolumn{4}{l}{\textit{With repulsion}} \\
Single particle & $29.92$ & $0.811$ & $0.196$ \\
Ensemble mean & $\mathbf{30.22}$ & $\mathbf{0.822}$ & $\mathbf{0.190}$ \\
\bottomrule
\end{tabular}
\end{table}

\end{document}